\documentclass[dvips]{article}

\usepackage{icrc2011}

\title{A Monte Carlo study to measure the energy spectra of the primary cosmic-ray components at the knee using a new Tibet AS core detector array}

\newcommand{\etal}{\MakeLowercase{\textit{et al. }}} 
\shorttitle{L.M. ZHAI \etal energy spectra of primary proton, helium
and iron}

\authors{M.~Amenomori$^{1}$, X.~J.~Bi$^{2}$, D.~Chen$^{3}$, W.~Y.~Chen$^{2}$, S.~W.~Cui$^{4}$,
Danzengluobu$^{5}$, L.~K.~Ding$^{2}$, X.~H.~Ding$^{5}$,
C.~F.~Feng$^{6}$, Zhaoyang Feng$^{2}$, Z.~Y.~Feng$^{7}$,
Q.~B.~Gou$^{2}$, H.~W.~Guo$^{5}$, Y.~Q.~Guo$^{2}$, H.~H.~He$^{2}$,
Z.~T.~He$^{4,2}$, K.~Hibino$^{8}$, N.~Hotta$^{9}$, Haibing~Hu$^{5}$,
H.~B.~Hu$^{2}$, J.~Huang$^{2}$, W.~J.~Li$^{2,7}$, H.~Y.~Jia$^{7}$,
L.~Jiang$^{2}$, F.~Kajino$^{10}$, K.~Kasahara$^{11}$,
Y.~Katayose$^{12}$, C.~Kato$^{13}$, K.~Kawata$^{3}$,
Labaciren$^{5}$, G.~M.~Le$^{2}$, A.~F.~Li$^{14,6,2}$, C.~Liu$^{2}$,
J.~S.~Liu$^{2}$, H.~Lu$^{2}$, X.~R.~Meng$^{5}$,
K.~Mizutani$^{11,15}$, K.~Munakata$^{13}$, H.~Nanjo$^{1}$,
M.~Nishizawa$^{16}$, M.~Ohnishi$^{3}$, I.~Ohta$^{17}$,
S.~Ozawa$^{11}$, X.~L.~Qian$^{6,2}$, X.~B.~Qu$^{2}$,
T.~Saito$^{18}$, T.~Y.~Saito$^{19}$, M.~Sakata$^{10}$,
T.~K.~Sako$^{12}$, J.~Shao$^{2,6}$, M.~Shibata$^{12}$,
A.~Shiomi$^{20}$, T.~Shirai$^{8}$, H.~Sugimoto$^{21}$,
M.~Takita$^{3}$, Y.~H.~Tan$^{2}$, N.~Tateyama$^{8}$,
S.~Torii$^{11}$, H.~Tsuchiya$^{22}$, S.~Udo$^{8}$, H.~Wang$^{2}$,
H.~R.~Wu$^{2}$, L.~Xue$^{6}$, Y.~Yamamoto$^{10}$, Z.~Yang$^{2}$,
S.~Yasue$^{23}$, A.~F.~Yuan$^{5}$, T.~Yuda$^{3}$, L.~M.~Zhai$^{2}$,
H.~M.~Zhang$^{2}$, J.~L.~Zhang$^{2}$, X.~Y.~Zhang$^{6}$,
Y.~Zhang$^{2}$, Yi~Zhang$^{2}$, Ying~Zhang$^{2}$,
Zhaxisangzhu$^{5}$, X.~X.~Zhou$^{7}$\ (The Tibet AS$\gamma$
Collaboration)}

\afiliations{$^{1}$Department of Physics, Hirosaki University, Hirosaki 036-8561, Japan\\
$^{2}$Key Laboratory of Particle Astrophysics, Institute of High
Energy Physics, Chinese
Academy of Sciences, Beijing 100049, China\\
$^{3}$Institute for Cosmic Ray Research, University of Tokyo, Kashiwa 277-8582, Japan\\
$^{4}$Department of Physics, Hebei Normal University, Shijiazhuang 050016, China\\
$^{5}$Department of Mathematics and Physics, Tibet University, Lhasa 850000, China\\
$^{6}$Department of Physics, Shandong University, Jinan 250100, China\\
$^{7}$Institute of Modern Physics, SouthWest Jiaotong University, Chengdu 610031, China\\
$^{8}$Faculty of Engineering, Kanagawa University, Yokohama 221-8686, Japan\\
$^{9}$Faculty of Education, Utsunomiya University, Utsunomiya 321-8505, Japan\\
$^{10}$Department of Physics, Konan University, Kobe 658-8501, Japan\\
$^{11}$Research Institute for Science and Engineering, Waseda
University, Tokyo 169-8555,
Japan\\
$^{12}$Faculty of Engineering, Yokohama National University, Yokohama 240-8501, Japan\\
$^{13}$Department of Physics, Shinshu University, Matsumoto 390-8621, Japan\\
$^{14}$School of Information Science and Engineering, Shandong
Agriculture University,
Taian 271018, China\\
$^{15}$Saitama University, Saitama 338-8570, Japan\\
$^{16}$National Institute of Informatics, Tokyo 101-8430, Japan\\
$^{17}$Sakushin Gakuin University, Utsunomiya 321-3295, Japan\\
$^{18}$Tokyo Metropolitan College of Industrial Technology, Tokyo 116-8523, Japan\\
$^{19}$Max-Planck-Institut f\"ur Physik, M\"unchen D-80805, Deutschland\\
$^{20}$College of Industrial Technology, Nihon University, Narashino 275-8576, Japan\\
$^{21}$Shonan Institute of Technology, Fujisawa 251-8511, Japan\\
$^{22}$RIKEN, Wako 351-0198, Japan\\
$^{23}$School of General Education, Shinshu University, Matsumoto
390-8621, Japan}

\abstract{A new hybrid experiment has been started by AS$\gamma$
experiment at Tibet, China, since August 2011, which consists of a
low threshold burst-detector-grid (YAC-II, Yangbajing Air shower
Core array), the Tibet air-shower array (Tibet-III) and a large
underground water Cherenkov muon detector (MD). In this paper, the
capability of the measurement of the chemical components (proton,
helium and iron) with use of the (Tibet-III+YAC-II) is investigated
by means of an extensive Monte Carlo simulation in which the
secondary particles are propagated through the (Tibet-III+YAC-II)
array and an artificial neural network (ANN) method is applied for
the primary mass separation. Our simulation shows that the new
installation is powerful to study the chemical compositions, in
particular, to obtain the primary energy spectrum of the major
component at the knee.}

\begin{document}
\maketitle

\section{Introduction}

$\;\;\;$The all-particle energy spectrum of primary cosmic rays is
well discriminated by a power law dN/dE $\propto$ E$^{-\gamma}$ over
many orders of magnitude, with $\gamma$ changes sharply from 2.7 to
3.1 at about 4 PeV [1]. The break of the all-particle energy
spectrum is called the "knee", and the corresponding energy range is
called "knee region". Although the existence of the knee is well
confirmed by experiments, there still is controversy on its origin.
In order to explain the sharpness of the knee, two scenarios [2]
(model A and model B) are proposed as shown in Fig.1 and Fig.2.

$\;\;\;\;$In model A, an excess component is assumed to overlap the
global component, and its spectrum shape suggests that it can be
attributed to nearby source(s) because it is surprisingly close to
the expected source spectrum of the diffuse shock acceleration.
Middle composition is predicted by this model at the knee. In model
B, a hard observed energy spectrum of each element from a given
source is assumed. The sharp knee can be explained by a
rigidity-dependent acceleration limit and the hard spectrum due to
nonlinear effects. Iron-dominant composition is predicted by this
model at the knee and beyond. In order to distinguish between Model
A and Model B and many other models, measurements of the chemical
composition around the knee, especially measurements of the primary
spectra of individual component till their knee will be essentially
important. Therefore, we have developed the Yangbajing Air shower
Core detector (YAC) which is operated along with Tibet air-shower
array (Tibet-III) and underground water Cherenkov muon detector
array (MD) simultaneously, as shown in Fig.3, and the second phase
of YAC is so called YAC-II. In this paper, the capability of the
measurement of the chemical components with use of the
(Tibet-III+YAC-II) is investigated. The simulation results using
(Tibet-III+YAC-II+MD) will be reported in the near further.

\begin{figure}[!t]
  \vspace{2mm}
  \centering
  \includegraphics[width=2.15 in]{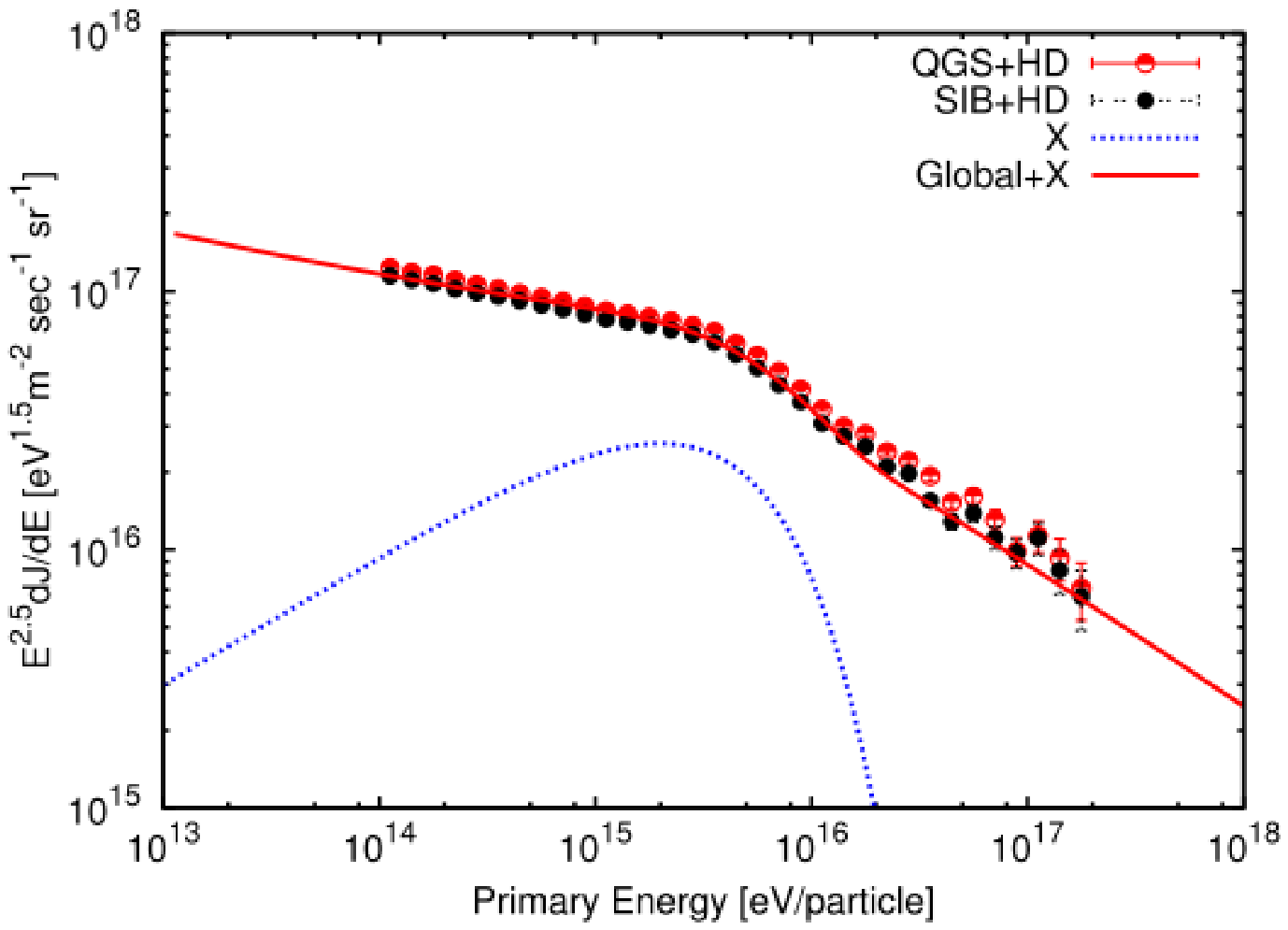}

  \caption{Model A: Sharp knee is attributed to extra component from nearby source.}
  \vspace{2mm}
  \centering
  \includegraphics[width=2.5 in]{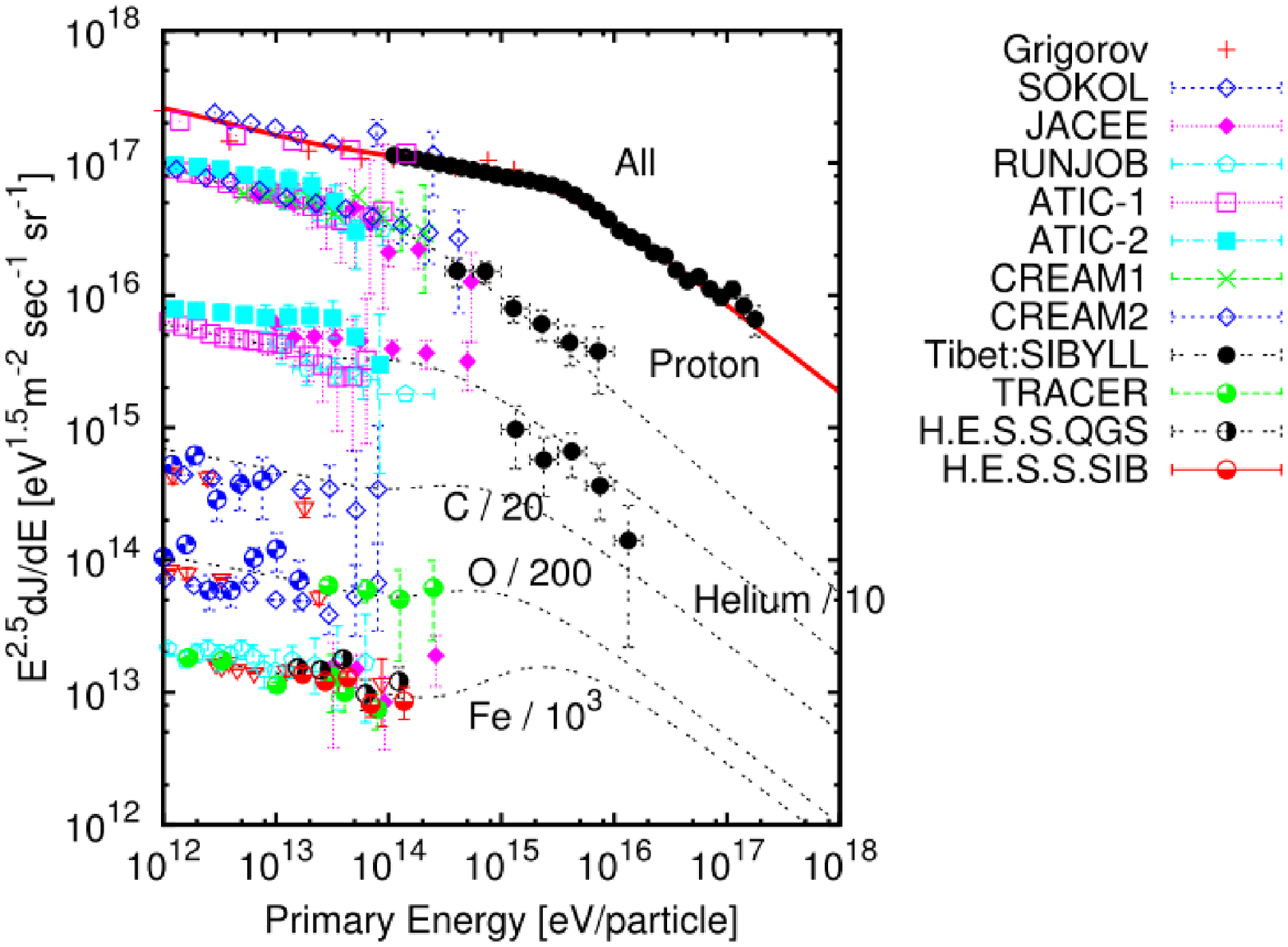}

  \caption{ Model B: Nonlinear effect in the
 diffusive shock acceleration mechanism can explain the structure of the knee when high
acceleration efficiency for heavy elements is assumed. }
  \label{double_fig}
 \end{figure}

\section{A New Tibet Hybrid Experiment}
$\;\;\;$ The Tibet new hybrid experiment (Tibet-III+YAC-II) has been
operated in Tibet, China, since August, 1st, 2011. The merit of this
experiment is that the atmosphere depth of the experimental site
(4300 m above sea level; 606 g/cm$^{2}$) is close to the maximum of
the air shower development with energies around the knee and the
shower maximum values are almost independent of the masses of
primary cosmic rays. The Tibet-III consists of 789 detectors while
the YAC-II consists of 124 detectors (as shown in Fig.3). The inner
100 plastic scintillator units of YAC-II are arranged as an array
(10$\times$10 grid) each with an area of 50 cm $\times $ 80 cm, with
1.875 m interval; and the outer 24 plastic scintillator units are
arranged around the inner array each with an area of 50 cm $\times$
100 cm. The outer 24 units are used to reject non core events whose
shower cores are far from the YAC-II array. Each detector of YAC-II
consists of lead plates with a thickness of 3.5 cm above the
scintillator to convert high energy electrons and gammas to
electromagnetic showers. Each unit of YAC-II is attached with two
photomultipliers of high-gain (HAMAMATSU: R4125) and low-gain
(HAMAMATSU: R5325) to cover the wide dynamic range from 1 MIP
(Minimum Ionization Particle) to 10$^{6}$ MIPs. The hardware of
YAC-II is described in [3].

\begin{figure}[!t]
  \vspace{2mm}
  \centering
  \includegraphics[width=3.2 in]{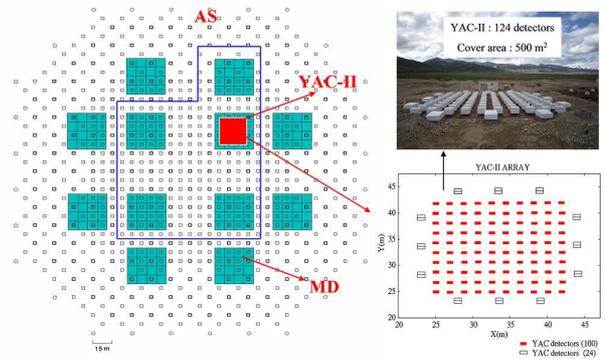}

  \caption{The Tibet-III+YAC-II+MD array. The Tibet-III consists of 789 detector units, the YAC-II consists of 124 detector units.
  The five MDs in
the blue frame are set up this year and acquiring data soon. }
  \label{single_fig}
 \end{figure}

$\;\;\;$The area of YAC-II array is about 500 m$^{2}$, locating near
the center of the Tibet-III, and operating simultaneously with it.
For an air shower event, the Tibet-III provides the arrival
direction ($\theta$), the air shower age ($Age$) and the air shower
size (${N_e}$) which is interrelated to primary energy, the YAC-II
measures the high energy electromagnetic particles in the core
region so as to obtain the characteristic parameters of air-shower
cores. When a YAC event is triggered, its accompanying air shower is
simultaneously recorded. The matching between an AS and a YAC event
is made by their arrival time stamps. The air-shower direction can
be estimated with an error smaller than 0.1$^{0}$ above 100 TeV, and
the primary energy resolution is estimated to be 12\% at energies
around 10$^{15}$ eV by our simulation.

\section{Simulation and Analysis}

$\;\;\;\;$A Monte Carlo (MC) simulation has been carried out on the
development of extensive air showers in the atmosphere and the
response in YAC-II and Tibet-III. The simulation code CORSIKA
(version 6.024) including QGSJET2 hadronic interaction models [4] is
used to generate AS events. Considering the aim of this simulation
is just to check the capability of the hybrid experiment, QGSJET2
interaction model and heavy dominated (HD) [1] primary composition
model are used. Primaries isotropically incident at the top of the
atmosphere within the zenith angles from 0 to 60 degrees are
injected into the atmosphere. The simulated data were analyzed in
the same manner as in the procedure for the experimental data
analysis. The electromagnetic showers in the lead layer induced by
electrons or photons that hit any detector unit of the array are
simulated by a subroutine which is based on the detector simulation
code EPICS [5]. The detector performance, the trigger efficiency of
detectors and the effective area are adequately taken into account
based on the experimental conditions. Normally, the following
quantities of YAC-II are used to characterize an air-shower core
event:

${N_b}$ - the number of shower particles under the lead plate of a
detector unit;

${N_{hit}}$ - the number of "fired" detector units with
${N_b}$$\geq$ a given threshold value;

${N_b}$$^{top}$ - the maximum
burst size among fired detectors;

$\sum$${N_b}$ - the total burst
size of all fired detector units;

$<R>$ - the mean lateral spread, $<R>$=$\sum$${r_i}$/($N_{hit}$-1);

$<${${N_b}$$R$}$>$ - the mean energy-flow spread,

 $<${${N_b}$$R$}$>$=$\sum$(${N_{bi}}$$\times$${r_i}$)/$N_{hit}$, where
${N_{bi}}$ and ${r_{i}}$ are the burst size in the $i^{th}$ fired
detector unit and the lateral distance from the air shower core to
the center of the $i^{th}$ fired detector, respectively.


$\;\;$ We divided the MC data into two datasets.  Due to the
difference of the detection efficiency, the first dataset selects
events that are enriched with ¡°proton+helium¡± origin (called
tagged-I dataset), while the second dataset contains events of
heavy-primary origin (called tagged-II dataset). The final select
conditions for tagged-I and tagged-II are as follows:

$\;\;$(1) ${N_b}$ $\geq$ 100, $N_{hit}$ $\geq$ 12, ${N_b}$$^{top}$
$\geq$ 3000, ${N_e}$ $\geq$ 1$\times$10$^5$;

 $\;\;$(2) ${N_b}$ $\geq$ 100,
$N_{hit}$ $\geq$ 20, ${N_b}$$^{top}$ $\geq$ 1500, ${N_e}$ $\geq$
5$\times$10$^5$.

$\;\;\;$ Besides, the detector unit with ${N_b}$$^{top}$ is
requested to be located at inner 8$\times$8 grid for both data-sets.
Then, we obtain 1.43$\times$10$^5$ and 5.17$\times$10$^4$ events for
the tagged-I and tagged-II dataset, respectively. The detection
efficiency S$\Omega$$A_{eff}$ is shown in Fig.4 .

\begin{figure}[!t]
  \vspace{0.5mm}
  \centering
  \includegraphics[width=2.55 in]{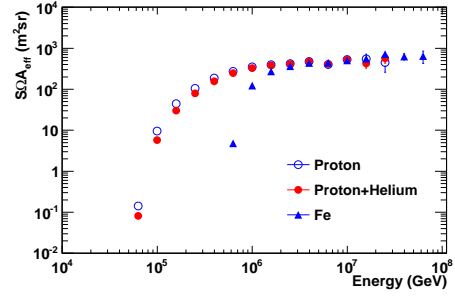}

  \caption{The detection efficiency S$\Omega$$A_{eff}$ of proton, proton+helium and iron.}
  \label{single_fig}
 \end{figure}

 $\;\;$ The separation of the primary mass is realized by use of a
 feed-forward artificial neural
network (ANN[6]) method, whose applicability to our experiment was
well confirmed by the MC simulation [7]. For one thing, we need to
separate protons from other nuclei by training the network with a
proton flag, and then separate proton+helium from other nuclei by
training the network with a proton+helium flag, thus we can get the
helium energy spectrum by subtracting the number of proton events
from the proton+helium events, so does iron. The following 8
parameters are input to the ANN with 40 hidden nodes and 1 output
unit:
(1)${N_{hit}}$, (2)${N_b}$$^{top}$, (3)$\sum$${N_b}$, (4)$<R>$,
(5)$<${${N_b}$$R$}$>$, (6)${N_e}$, (7)$\theta$, (8)$Age$, where the
first five parameters are given by YAC-II, and the last three are
obtained by Tibet-III.

\section{Results and Discussion}

$\;\;\;$The ANN training results of proton, proton+helium and iron
are presented in Fig.5, where average purity and selection
efficiency over whole energy range are shown. The ANN output value
$T$ is used to separate the primary nuclei groups. In this paper,
the events with $T$ $\leq$ ${T_c}$ (or $T$ $\geq$ ${T_c}$ ) are
regarded as proton or proton+helium group (or iron group). The value
${T_c}$, purity (p) and selection rate ($\varepsilon$) of events
satisfying the criterion at various energy regions are summarized in
Table 1.

\begin{table*}[t]
\caption{The purity (p)(\%) and the selection rate
($\varepsilon$)(\%) of the selected primary groups. }
\begin{center}
\begin{tabular}{|c|c|c|c|c|c|c|c|c|c|}

\hline\hline
\multicolumn{2}{|c|}{Primary Energy } & \multicolumn{2}{|c|}{$10^{14}-10^{15}$ eV } & \multicolumn{2}{|c|}{$10^{15}-10^{16}$ eV } & \multicolumn{2}{|c|}{$10^{16}-10^{17}$ eV } \\
\hline
       & $T_c(Mod)$ &$\;\;$ p $\;\;$     & $\varepsilon$      & $\;\;$p $\;\;$    & $\varepsilon$       & $\;\;$p $\;\;$    & $\varepsilon$    \\
\hline
P          & 0.2    & 87.6 $\pm$ 1.1     & 45.5 $\pm$ 0.5     & 77.3 $\pm$ 5.9     & 19.5 $\pm$ 1.2     &                   &                  \\
\hline
P+He       & 0.1    & 96.7 $\pm$ 0.8     & 80.8 $\pm$ 0.6     & 87.3 $\pm$ 3.5     & 39.1 $\pm$ 1.3     &                   &                  \\
\hline
Fe         & 0.7    &                    &                    & 81.6 $\pm$ 1.6     & 55.3 $\pm$ 1.0     & 81.6 $\pm$ 8.5    & 69.4 $\pm$ 7.0   \\

\hline
\end{tabular}

\label{table_single}
\end{center}
\end{table*}

\begin{figure}[!t]
  \vspace{1mm}
  \centering
  \includegraphics[width=2.1 in]{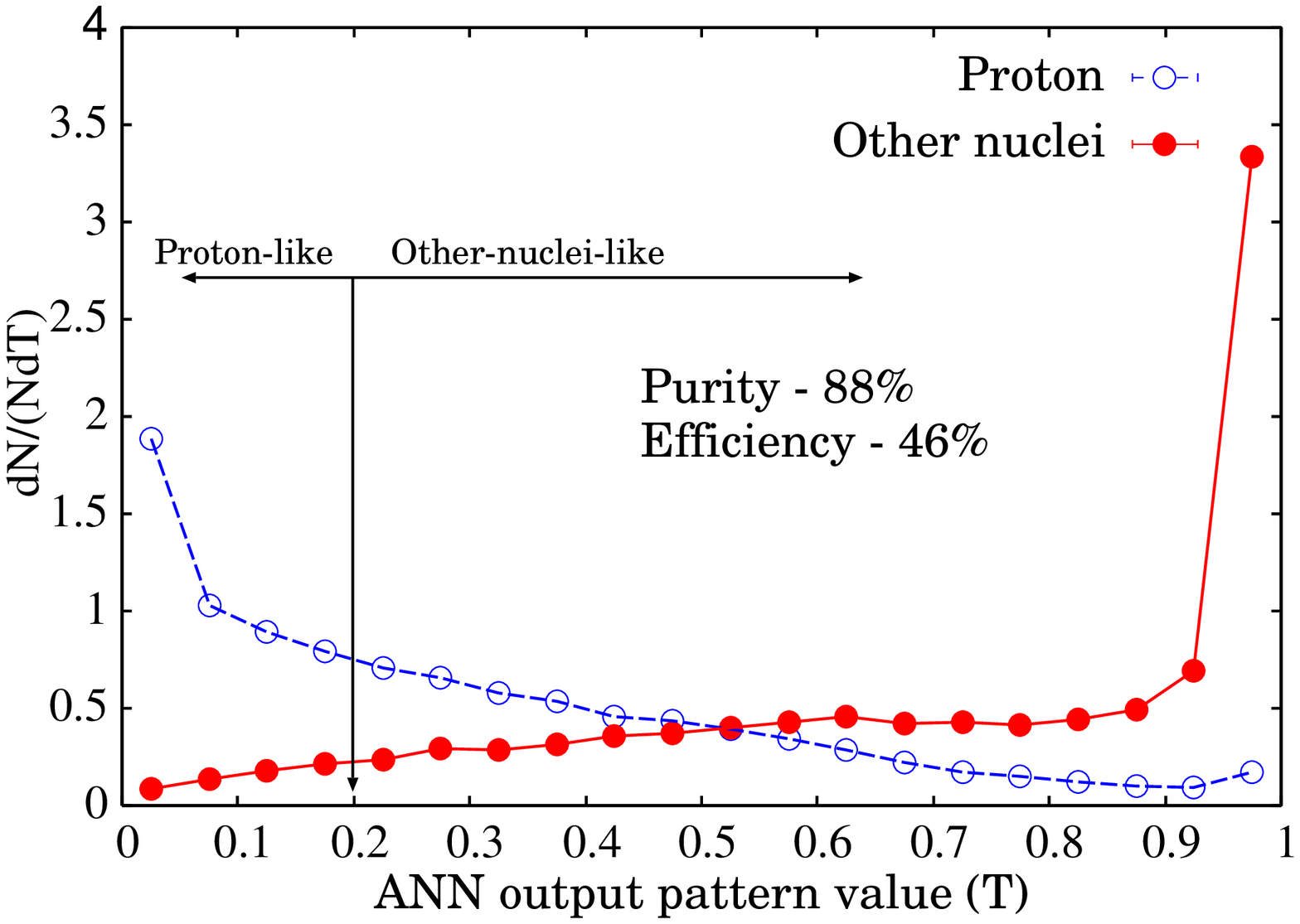}
  \includegraphics[width=2.1 in]{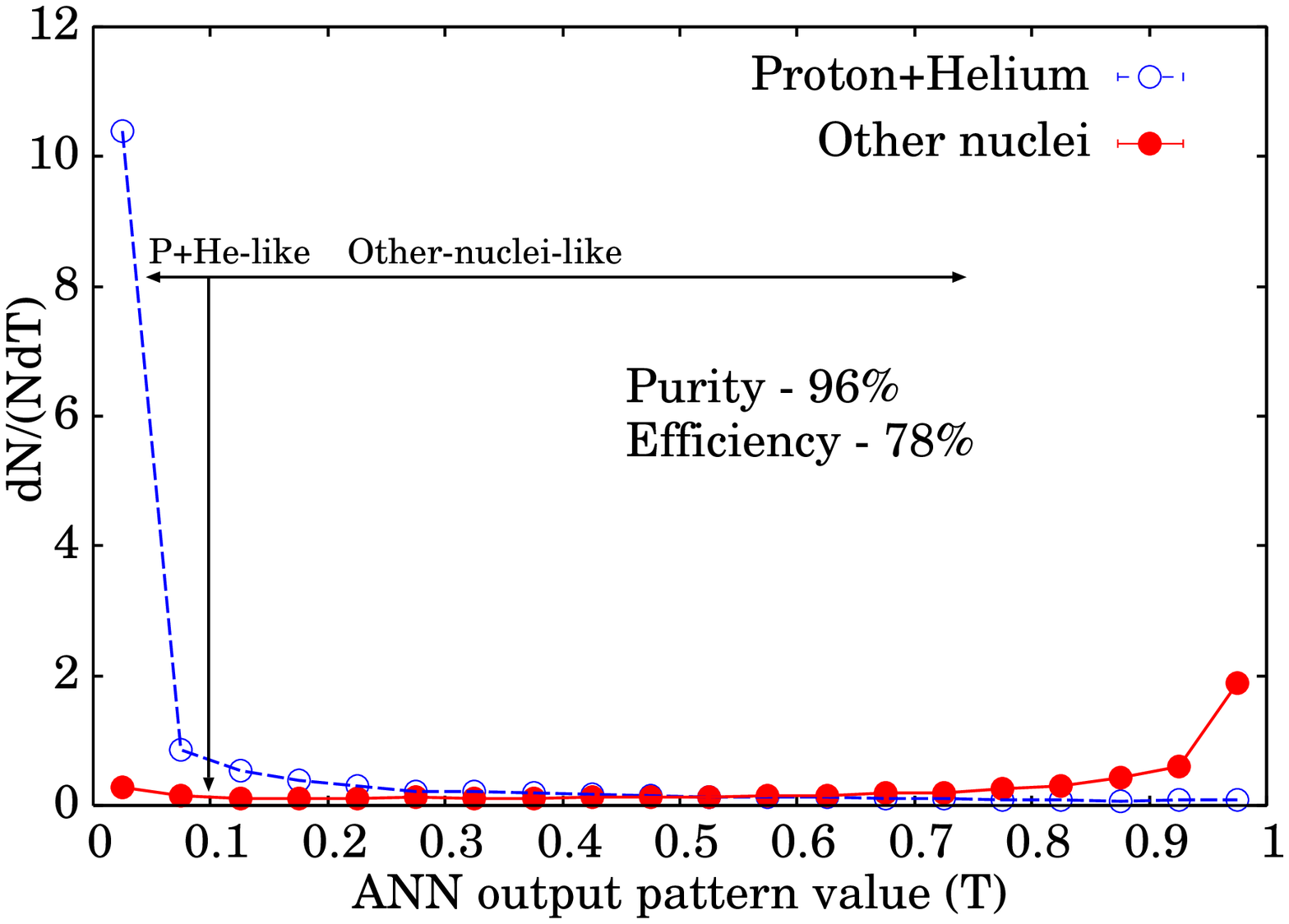}
  \includegraphics[width=2.1 in]{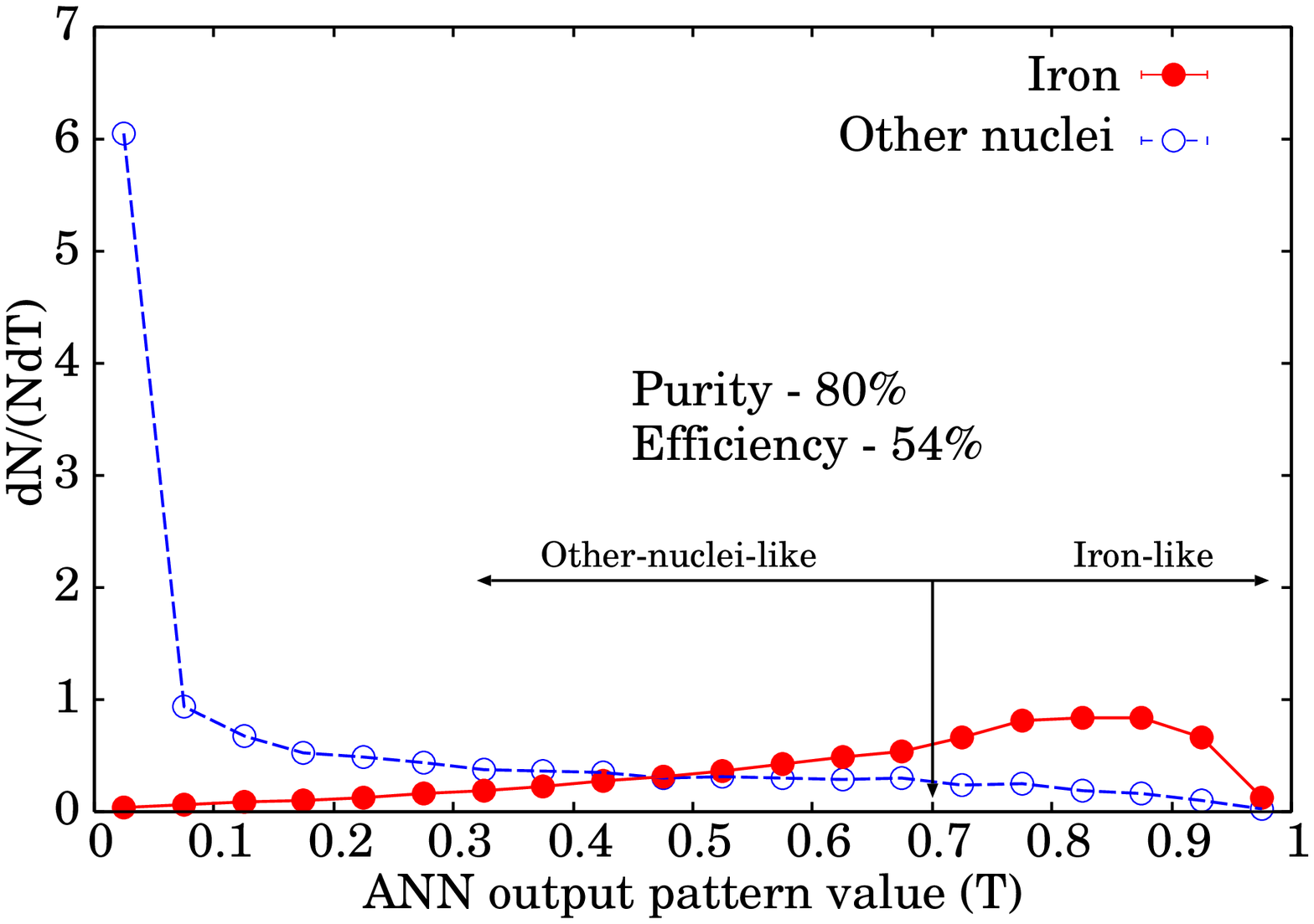}

  \caption{The ANN training results of proton, proton+helium and iron. The average purity and selection rate over whole energy range of protons are 88\%, 46\%
at ${T_c}$ = 0.2, while 96\%, 28\% for proton+helium at ${T_c}$ =
0.1, 80\%, 54\% for iron at ${T_c}$ = 0.7. }
  \label{double_fig}
 \end{figure}

$\;\;$ The Fig.6 shows the estimated primary energy spectra of
proton, helium and iron compared with the assumed ones. One can see
that the assumed primary energy spectrum of proton, helium and iron
are well reproduced by the estimated ones respectively, and the
estimated ones could well connect with the results obtained by
direct observation, as shown in Fig.7. It needs to be remarked that
the iron spectra seem to be higher than the observed ones, just
because of the HD model we used. The results above show that the new
burst detector array is powerful to study the chemical composition,
in particular, to obtain the primary energy spectrum of the major
component at the knee.

\begin{figure}[!t]
  \vspace{1mm}
  \centering
  \includegraphics[width=2.6 in]{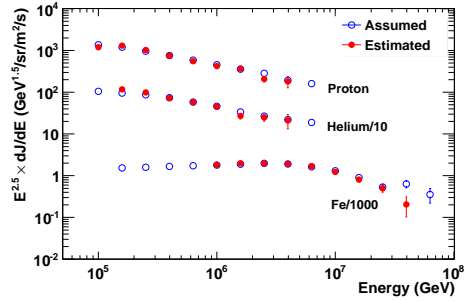}

  \caption{Energy spectra of primary proton, helium and iron.}
  \label{single_fig}
 \end{figure}

 \begin{figure}[!t]
  \vspace{1mm}
  \centering
  \includegraphics[width=2.6 in]{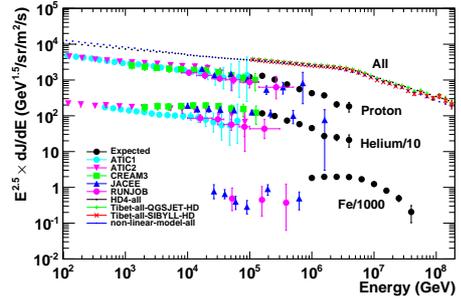}

  \caption{The expected spectra
of proton, helium and iron obtained by MC compared with other
experimental data.}
  \label{single_fig}
 \end{figure}

\section{Acknowledgements}
$\;\;$This work is supported by the Chinese Academy of Sciences
(H9291450S3) and  the  Key  Laboratory of  Particle  Astrophysics,
Institute of High Energy Physics, CAS.  The Knowledge Innovation
Fund (H95451D0U2 and H8515530U1) of IHEP, China and the project
Y0293900TF of NSFC also provide support to this study.



\clearpage


\begin{thebibliography}{}

\bibitem{1} M. Amenomori et al., ApJ., 678:1165-1179, 2008.


\bibitem{2} M. Shibata, Y. Katayose, J. Huang and  D. Chen, ApJ., 716,
1076¨C1083, 2010.


\bibitem{3} M. Amenomori et al., ICRC 32 (HE 1.4 ID: 1241), 2011.

\bibitem{4} D. Heck et al., Report FZKA 6019, 1998;
J. Knapp, D. Heck et al., Report FZKA 3640, 1997; D. Heck et al.,
Report FZKA 5828, Forshungszentru Karldruhe, 1996. Available from
http://www-ik3.fzk.deheck/corsika
/physics$_{-}$description/corsika$_{-}$phys.html.


\bibitem{5} K. Kasahara et al., http://eweb.b6.kanagawa-u.ac.jp/
~Kasahara/ResearchHome/EPICSHome/Index.html.

\bibitem{6} L. Lonnblad et al., Comp. Phys. Com. 81, 185 (1994).

\bibitem{7} M. Amenomori et al., Physics Letters B, 632, 58-64, 2006.



\end{thebibliography}
\end{document}